\documentclass[prl,aps,twocolumn,superscriptaddress,floatfix]{revtex4}
  
\usepackage[latin1]{inputenc} 
\usepackage{graphicx}  
\DeclareMathAlphabet{\mathpzc}{OT1}{pzc}{m}{it}
\begin{document} 
 
\title{Quantum and classical criticality in a dimerized quantum 
antiferromagnet} 

\author{P. Merchant}
\affiliation{London Centre for Nanotechnology and Department of Physics and 
Astronomy, University College London, London WC1E 6BT, UK} 
\author{B. Normand}
\affiliation{Department of Physics, Renmin University of China, Beijing
100872, China}
\author{K. W. Kr\"amer}
\affiliation{Department of Chemistry and Biochemistry, University of Bern, 
CH--3012 Bern 9, Switzerland}
\author{M. Boehm}
\affiliation{Institut Laue Langevin, BP 156, 38042 Grenoble Cedex 9, France}
\author{D. F. McMorrow}
\affiliation{London Centre for Nanotechnology and Department of Physics and 
Astronomy, University College London, London WC1E 6BT, UK} 
\author{Ch. R\"uegg}
\affiliation{London Centre for Nanotechnology and Department of Physics and 
Astronomy, University College London, London WC1E 6BT, UK} 
\affiliation{Laboratory for Neutron Scattering, Paul Scherrer Institute, 
CH--5232 Villigen, Switzerland}
\affiliation{DPMC--MaNEP, University of Geneva, CH--1211 Geneva, Switzerland}
 
\date{\today} 
 
\maketitle 

{\bf A quantum critical point (QCP) is a singularity in the phase diagram 
arising due to quantum mechanical fluctuations. The exotic properties of 
some of the most enigmatic physical systems, including unconventional metals 
and superconductors, quantum magnets, and ultracold atomic condensates, have 
been related to the importance of the critical quantum and thermal 
fluctuations near such a point. However, direct and continuous control of 
these fluctuations has been difficult to realize, and complete thermodynamic 
and spectroscopic information is required to disentangle the effects of 
quantum and classical physics around a QCP. Here we achieve this control 
in a high-pressure, high-resolution neutron scattering experiment on the 
quantum dimer material TlCuCl$_3$. By measuring the magnetic excitation 
spectrum across the entire quantum critical phase diagram, we illustrate 
the similarities between quantum and thermal melting of magnetic order.
We prove the critical nature of the unconventional longitudinal (``Higgs'') 
mode of the ordered phase by damping it thermally. We demonstrate the 
development of two types of criticality, quantum and classical, and use 
their static and dynamic scaling properties to conclude that quantum and 
thermal fluctuations can behave largely independently near a QCP.}

In ``classical'' isotropic antiferromagnets, the excitations of the ordered 
phase are gapless spin waves emerging on the spontaneous breaking of a 
continuous symmetry \cite{Goldstone}. The classical phase transition, 
occurring at the critical temperature $T_N$, is driven by thermal 
fluctuations. In quantum antiferromagnets, quantum fluctuations suppress 
long-range order, and can destroy it completely even at zero temperature 
\cite{rchn}. The ordered and disordered phases are separated by a quantum 
critical point (QCP), where quantum fluctuations restore the broken symmetry 
and all excitations become gapped, giving them characteristics fundamentally 
different from the Goldstone modes on the other side of the QCP (Fig.~1). 
At finite temperatures around a QCP, the combined effects of quantum and 
thermal fluctuations bring about a regime where the characteristic energy 
scale of spin excitations is the temperature itself, and this quantum 
critical (QC) regime has many special properties \cite{rs}. 

\begin{figure}[t]
\includegraphics[width=8.5cm]{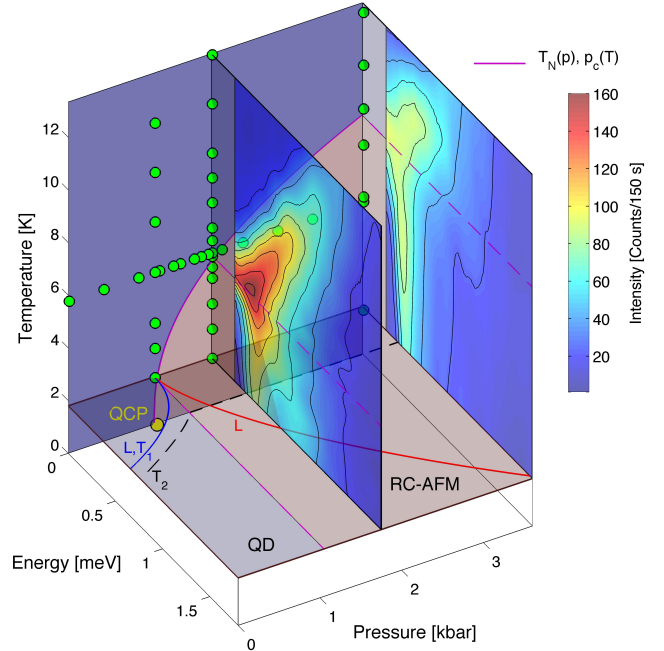}
\caption{Pressure-temperature phase diagram of TlCuCl$_3$ extended to finite 
energies, revealing quantum and thermal critical dynamics. The rear panel is 
the bare $(p,T)$ phase diagram at energy $E = 0$ meV, in which the magenta 
line shows the N\'eel temperature as a function of pressure, $T_N (p)$
\protect{\cite{rrfsskgm}}, and the green points depict the temperature and 
pressure values studied. Full details of this panel are presented in Fig.~4(c).
The centre $(E,T)$ panel shows neutron intensity data collected from $T = 1.8$ 
K to 12.7 K at $p = 1.75$ kbar, where $T_N = 5.8$ K. The rightmost $(E,T)$ 
panel shows the corresponding data at $p = 3.6$ kbar, where $T_N = 9.2$ K. 
The data in both $(E,T)$ panels display a clear softening of the magnetic 
excitations at $T_N (p)$. The bottom $(p,E)$ panel indicates the softening 
of the excitations, measured at $T = 1.8$ K, across the QPT 
\protect{\cite{rrnmfmkggmb}}. T$_1$, T$_2$ and L denote the three gapped 
triplet excitations of the quantum disordered (QD) phase. In the renormalized 
classical antiferromagnetic (RC-AFM) ordered phase, these become respectively 
the gapless Goldstone mode, which is a transverse spin wave, a gapped 
(anisotropic) spin wave, and the longitudinal ``Higgs'' mode (see text).} 
\label{fig1}  
\end{figure} 

Physical systems do not often allow the free tuning of a quantum fluctuation 
parameter through a QCP. The QC regime has been studied in some detail in 
heavy-fermion metals with different dopings, where the quantum phase 
transition (QPT) is from itinerant magnetic phases to unusual metallic 
or superconducting ones \cite{rlrvw,rkrlf,rsafgjklssss}, in organic materials 
where a host of insulating magnetic phases become (super)conducting 
\cite{rbj,rkk}, and in cold atomic gases tuned from superfluid to 
Mott-insulating states \cite{rbdz,rzhtc}. However, the dimerized quantum 
spin system TlCuCl$_3$ occupies a very special position in the experimental 
study of QPTs. The quantum disordered phase at ambient pressure and zero 
field has a small gap to spin excitations. An applied magnetic field closes 
this gap, driving a QPT to an ordered phase, a magnon condensate in the 
Bose-Einstein universality class, with a single, nearly massless excitation 
\cite{rnoot,rgrt}. 

Far more remarkably, an applied pressure also drives a QPT to an ordered 
phase \cite{rtgfou}, occurring at the very low critical pressure $p_c = 1.07$ 
kbar \cite{rrfsskgm} and sparking detailed studies \cite{rmnrs,rjs}. This 
ordered phase is a different type of condensate, whose defining feature 
is a massive excitation, a ``Higgs boson'' or longitudinal fluctuation 
mode of the weakly ordered moment \cite{raw,rltn}. This excitation, which 
exists alongside the two transverse (Goldstone) modes of a conventional 
well-ordered magnet, has been characterized in detail by neutron spectroscopy 
with continuous pressure control through the QPT \cite{rrnmfmkggmb} and 
subsequently by different theoretical approaches \cite{rpaa,roks}. TlCuCl$_3$ 
is therefore an excellent system for answering fundamental questions about 
the development of criticality, the nature of the QC regime, and the interplay 
of quantum and thermal fluctuations by controlling both the pressure and the 
temperature.

Here we present inelastic neutron scattering (INS) results which map 
the evolution of the spin dynamics of TlCuCl$_3$ throughout the quantum 
critical phase diagram in pressure and temperature. The spin excitations 
we measure exhibit different forms of dynamical scaling behaviour arising 
from the combined effects of quantum and thermal fluctuations, particularly 
on crossing the QC regime and at the line of phase transitions to magnetic 
order (Fig.~1). To probe these regions, we collected spectra up to 1.8 meV 
for temperatures between $T = 1.8$ K and 12.7 K, and over a range of 
pressures. Our measurements were performed primarily at $p = 1.05$ kbar 
($\simeq p_c$ at the lowest temperatures), 1.75 kbar, and 3.6 kbar, and 
also for all pressures at $T = 5.8$ K. Most measurements were made at the 
ordering wavevector, ${\bf Q}_0 = (0 \; 4 \; 0)$ reciprocal lattice units 
(r.l.u.), and so concern triplet mode gaps. From the INS selection rules, 
only one transverse mode of the ordered phase is observable at ${\bf Q} = 
{\bf Q}_0$, and it is gapped ($\Delta_a = 0.38$ meV) due to a 1\% exchange 
anisotropy \cite{rrnmfmkggmb}. In the summary presented in Fig.~1, the 
contours represent scattered intensities at two selected pressures $p > p_c$. 
Both panels show strong QC scattering and a nontrivial evolution of the mode 
gaps and spectral weights with both $p$ and $T$, which is quantified in Fig.~2.

\begin{figure}[t] 
\begin{center}
\includegraphics[width=8.5cm]{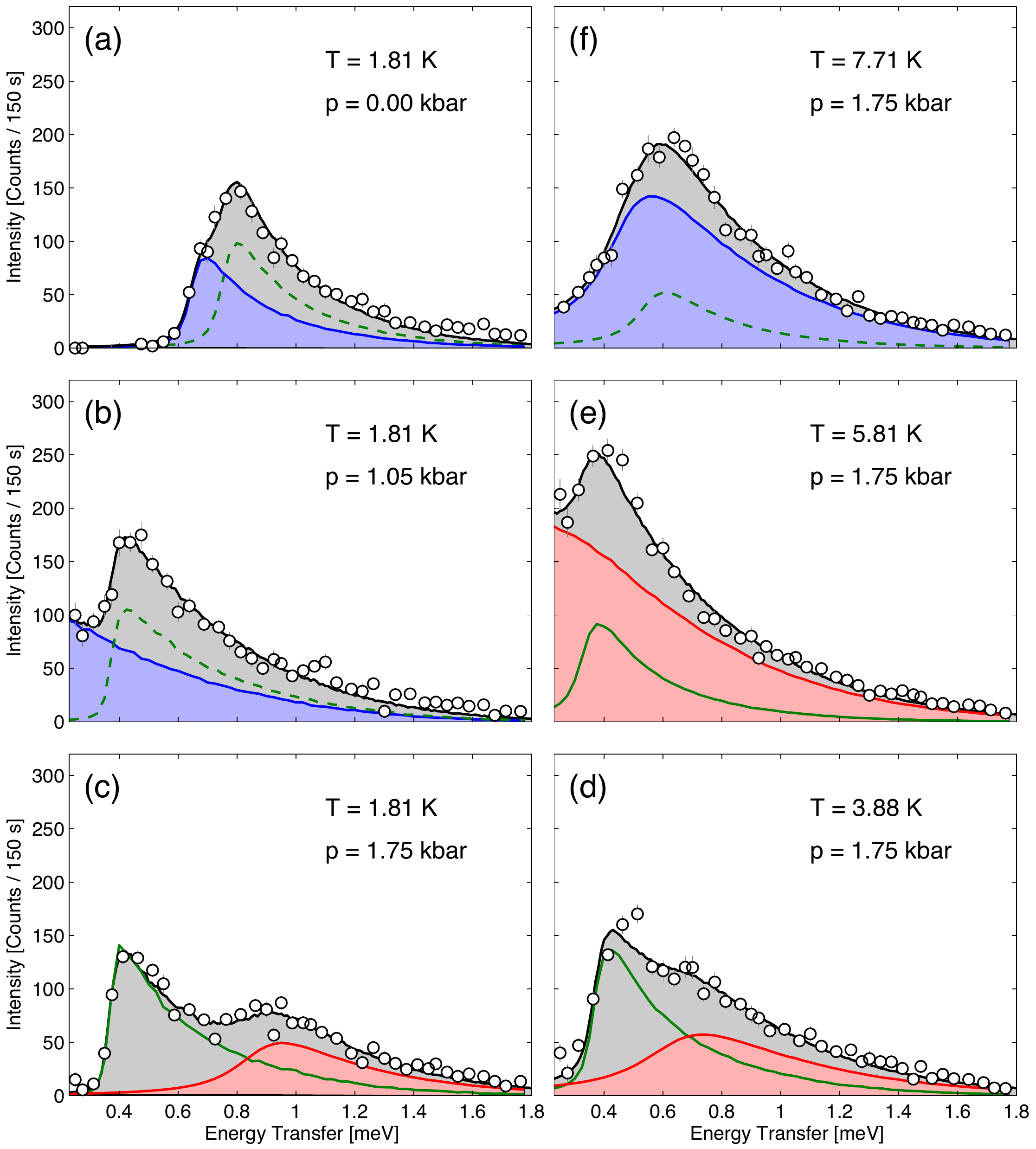}
\end{center}
\caption{INS spectra collected at ${\bf Q}$ = (0 4 0) r.l.u., the wavevector 
of the minimum energy gap in the QD phase, where magnetic order is induced 
with increasing pressure. (a-c) Evolution of triplet excitations during 
pressure-induced magnetic ordering, as the lower mode (L and T$_1$, blue) 
changes continuously into the longitudinal mode (L, red), while the 
anisotropic mode (T$_2$, green) remains gapped. (d-f) Evolution as increasing 
temperature lowers the longitudinal-mode gap to zero at $T = 5.8$ K, above 
which all three modes are gapped in a disordered (QC) state. }
\label{fig2}  
\end{figure}

Figures 2(a-c) show respectively the measured intensities for pressures 
below, at, and above the QPT at a fixed low temperature. Fits to the 
lineshapes of the separate excitations were made by a resolution 
deconvolution requiring both the gap and the local curvature of the 
mode dispersion, which was taken from a finite-temperature bond-operator 
description \cite{rrmnmfkgbsm,rnru}. The distinct contributions from 
transverse and longitudinal fluctuations change position systematically 
as the applied pressure induces magnetic order. The intensity of the 
longitudinal mode is highlighted in red in Fig.~2(c). Figures 2(d-f) 
show respectively the measured intensities for temperatures below, at, 
and above the phase transition [$T_N(p)$] at a fixed pressure $p > p_c$. 
Quantitatively, the intensity and the linewidth increase from the left 
to the right panels due to the temperature. Qualitatively, the thermal 
evolution is almost exactly analogous to a change in the pressure, with 
the spectral weight of the longitudinal mode softening at $T_N (p = 
1.75~{\rm kbar}) = 5.8$ K but moving again to finite energies at 
temperatures above $T_N$. 

\begin{figure}[t] 
\begin{center}
\includegraphics[width=8.5cm]{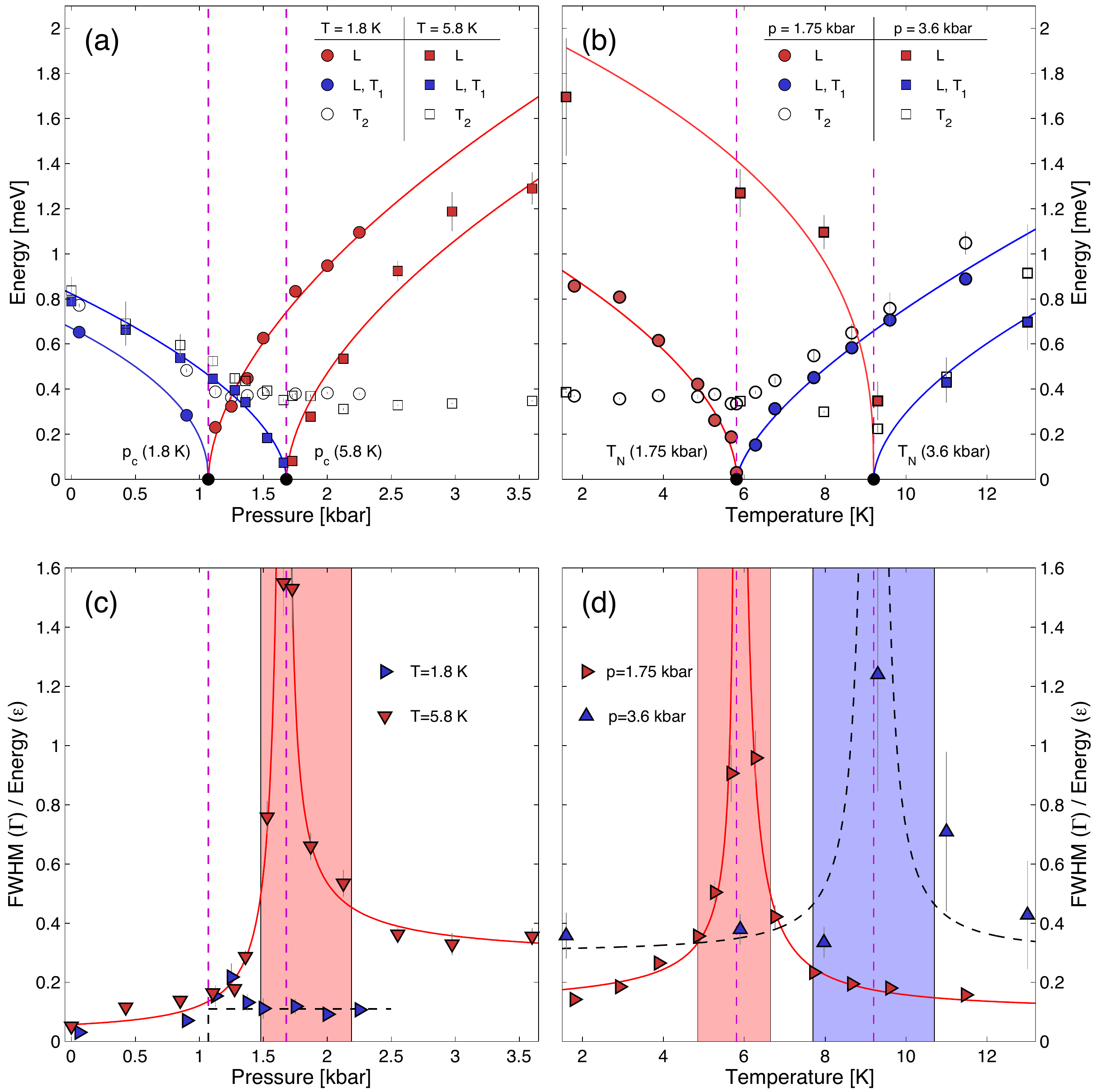}
\end{center}
\caption{Spin dynamics at the quantum and thermal ``melting'' transitions. 
(a) Quantum melting of magnetic order, shown by the triplet gaps for two 
different temperatures ($T = 1.8$ and 5.8 K), occurs from right to left. 
(b) Thermal melting, shown for two different pressures ($p = 1.75$ kbar 
and 3.6 kbar), occurs from left to right. Open black circles and squares 
give the energies of the anisotropic transverse excitations (T$_2$), which 
remain gapped, while filled circles and open squares show the longitudinal 
mode (red in the ordered phase, blue in the disordered one). The lines 
are power-law fits, described in the text. (c,d) Linewidth-to-energy ratios 
$\Gamma_{\bf Q}/\epsilon_{\bf Q}$ for the longitudinal mode across the phase 
transition as a function of pressure at $T = 1.8$ and 5.8 K (c) and of 
temperature at $p = 1.75$ and 3.6 kbar (d); lines are guides to the eye 
and shaded regions are explained in the text.}
\label{fig3}  
\end{figure}

We analyze these results in detail by extracting the excitation energies 
$\epsilon_{\bf Q}$ and linewidths $\Gamma_{\bf Q}$ from the data of Fig.~2. 
Figure 3(a) shows the evolution of the mode gaps ($\epsilon_{{\bf Q} = {\bf Q}_0}$) 
with pressure for $T = 1.8$ K (the QPT, {\it cf.}~Ref.~\cite{rrnmfmkggmb}) 
and $T = 5.8$ K. The longitudinal mode of the ordered phase appears on the 
right, and softens with decreasing pressure until $p_c(T)$. The Goldstone mode 
is not visible due to the scattering geometry. At pressures below $p_c(T)$, 
the effect of dimer-based quantum fluctuations is to destroy the magnetic 
order and gap all the modes. The lines are best fits to power laws of the 
form $\Delta (p) = A |p - p_c|^{\gamma_p}$, which we discuss below.

Figure 3(b) shows the evolution of the mode gaps over the temperature range 
$1.8$ K $< T < 12.7$ K for pressures $p = 1.75$ kbar and 3.6 kbar. Here the 
ordered phase is on the left, where the longitudinal mode, which dominates 
the low-energy scattering around the critical point, becomes soft at a N\'eel 
temperature $T_N(p)$ determined by the pressure. They reemerge on the right  
as gapped triplets of the thermally disordered QC phase. The lines in this 
figure are best fits to power laws of the form $\Delta (T) = B |T -
 T_N|^{\gamma_T}$. The similarity between quantum and thermal melting shown 
in Figs.~2 and 3 is a remarkable result. It is essential to note that the 
disorder in Fig.~3(b) is thermal, and not due to quantum fluctuations. 
Thermal fluctuations in a quantum dimer system, whose triplet excitations 
are hard-core bosons, do not simply broaden and damp the modes of the ordered 
magnet, but cause a very specific and systematic evolution of the spectral 
weight \cite{rrmnmfkgbsm}. On the ordered side, the massive, longitudinal 
mode becomes gapless at the classical phase transition, while on the 
disordered side there is not merely a featureless paramagnet but a clear 
gapped excitation. This is also the case for the pressure-driven transition 
at finite temperatures [Fig.~3(a)], where the symmetry is restored before 
all three excitations become gapped modes of the QC phase. 

Figures 3(c) and (d) show the linewidths of the longitudinal mode, 
measured respectively through the quantum and thermal transitions, 
in the form of the ratio $\Gamma_{\bf Q} / \epsilon_{\bf Q} = \alpha_L$. For 
the pressure-induced phase transition in Fig.~3(c), the ratio vanishes at 
$T = 1.8$ K for the well-defined excitations on the disordered side and 
remains constant (with $\alpha_L \simeq 0.15$) on the ordered side of the 
QCP, demonstrating its critically damped nature \cite{rrnmfmkggmb}. However, 
this is not at all the case at $T = 5.8$ K, where the divergence of $\alpha_L 
(T)$ shows the longitudinal mode becoming overdamped in the presence of thermal 
fluctuations. For the thermally driven phase transition at $p = 1.75$ and 
3.6 kbar [Fig.~3(d)], the ratio also diverges on approaching the critical 
temperature $T_N(p)$. 

The QC regime is the area around the line $p = p_c$ where the intrinsic 
energy scale of the system (the gap $\Delta$ in the QD phase, or $T_N$ in 
the ordered phase \cite{rchn}) is lower than the temperature \cite{rs}.
Near $p_c$, the measured neutron intensities [Fig.~4(a)] show a broad range 
over which spin excitations are present, with a peak along a line corresponding 
approximately to $\hbar \epsilon_{\bf Q} = k_{\rm B}T$. This ``$\omega/T$'' 
scaling property \cite{rs} is evident in the self-similar nature of the 
spectra at different temperatures. The QCP is the point where the intrinsic 
energy scale vanishes, and thus states become available at all energies; it 
is the maximum in their occupation that scales with $T$, and hence the 
temperature becomes the new characteristic energy scale. The microscopic 
origin of this ``thermal gap'' in the measured spectrum is mutual blocking 
of the hard-core triplet excitations \cite{rrmnmfkgbsm}. As shown in Fig.~4(b), 
the linewidth $\Gamma_{\bf Q} = \alpha_c \epsilon_{\bf Q}$ ($\alpha_c \simeq 0.14$) 
also scales linearly with $T$, illustrating that critical damping is an 
essential property of QC excitations. 

\begin{figure}[t] 
\begin{center}
\includegraphics[width=8.5cm]{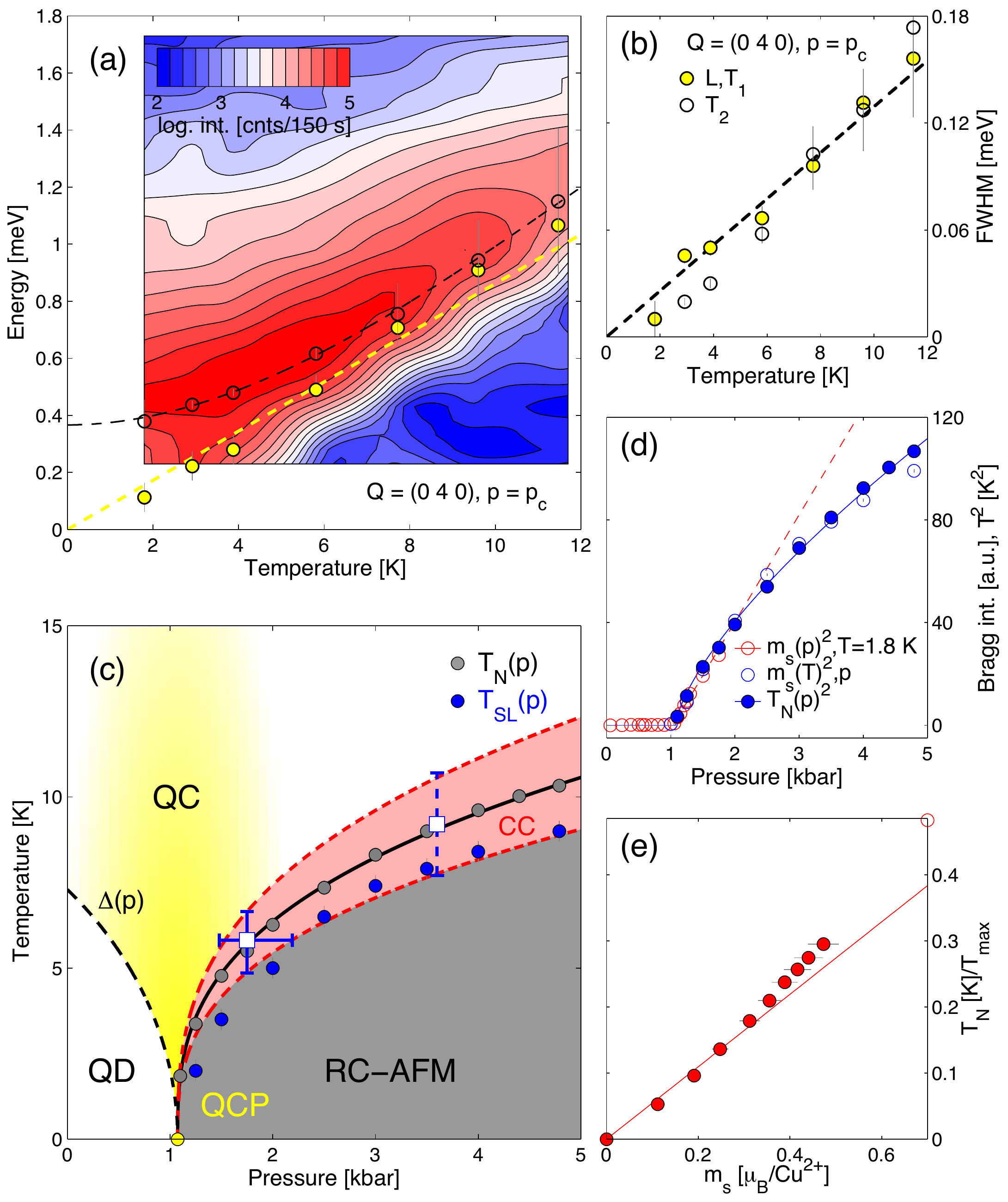}
\end{center}
\caption{Quantum and classical criticality. (a) Scattered neutron intensity 
at $p = p_c$ as a function of temperature. Points show the energies 
$\epsilon_{\bf Q}$ extracted from the intensity for the modes becoming 
gapless (L and T$_1$, yellow) and gapped (T$_2$, black) as $T \rightarrow 0$. 
(b) $\Gamma_{\bf Q}$ as a function of $T$ at $p = p_c$. (c) Complete experimental 
phase diagram, showing quantum disordered (QD), quantum critical (QC), 
classical critical (CC), and renormalized classical (RC-AFM) phases. The 
dashed lines denote energy scales marking crossovers in behaviour. Grey 
symbols denote $T_N(p)$ \cite{rrfsskgm}, blue symbols labelled $T_{SL}(p)$ 
show the limit of classical critical scaling in the data for the staggered 
magnetization, $m_s(T)$, and the blue bars are taken from $\Gamma_{\bf Q}/ 
\epsilon_{\bf Q} (T)$ (see text). (d) Linear proportionality of the measured 
$T_N(p)$ and $m_s(p)$ \cite{rrfsskgm}. (e) Scaling of $T_N$ and $m_s$, 
including one high-$p$ data point taken from Ref.~\cite{rokontt} for an 
absolute calibration of the ordered moment. Red lines in panels (d) and (e) 
represent scaling behaviour discussed in the text.} 
\label{fig4}  
\end{figure}

The experimental phase diagram is shown in Fig.~4(c), and contains all four 
regions characteristic of the QCP \cite{rs,rlrvw}. Between the quantum 
disordered and renormalized classically ordered regimes \cite{rchn}, the 
dominant behaviour is QC ($\omega/T$) scaling. On the line of classical phase 
transitions, the intrinsic energy scale is $T_N(p)$ but the excitation energy 
is driven to zero. This results in the properties, in particular the static 
scaling relations, of a classical critical (CC) regime. We show below that 
the scaling exponents in the QC and CC regions have approximately the values 
expected theoretically. However, for a real system such as TlCuCl$_3$, they 
can differ over broad crossover regions determined by the relative size of 
the intrinsic and excitation energy scales, and departures from universal 
behaviour may also arise due to microscopic details of the Hamiltonian.

We begin the analysis of our results not with static exponents but with the 
dynamical ones extracted from the power-law fits to the gaps in Figs.~3(a)
and (b). At the pressure-controlled transition, all of the exponents we 
measure fall in the range $0.46(6) \le \gamma_p \le 0.57(6)$, which within 
experimental error is $\gamma_p = 1/2$. This is the mean-field expectation 
in the generic case that the $p$-dependence of the exchange parameters 
$J_{ij}(p)$ is predominantly linear \cite{rrnmfmkggmb}. It applies both for 
the disordered phase ($\Delta_{QD}$) and for the gap ($\Delta_L$) of the 
longitudinal mode, with a multiplicative prefactor $\Delta_L = \sqrt{2} 
\Delta_{\rm QD}$ at $T = 0$ \cite{rssc}. At higher temperatures we find no 
strong departures from universality in this exponent. However, fits including 
data points at the higher energies are generally less reliable, suggesting 
that these begin to depart from the universal regime. 

For the classical (temperature-controlled) phase transition, the expectation 
for the QC phase is $\Delta \propto \xi^{-1} \propto (T - T_N)^\nu$, where $\nu 
\simeq 0.67$ for an XY spin symmetry and 0.70 for SU(2) symmetry \cite{rzj}. 
The 1\% exchange anisotropy in TlCuCl$_3$ reduces the spin symmetry from SU(2) 
to XY below energy scales of order 5 K. Our data for $\gamma_T$ at $p = 1.75$ 
kbar give $\gamma_T = 0.67(9)$, in excellent agreement with these exponents; 
at $p = 3.6$ kbar we do not have sufficient data for a reliable fit. At $p = 
p_c$, however, we find the expected QC scaling form $\gamma_T = 1$, and the 
width of the crossover regime remains an open question. Deducing a scaling 
exponent $\gamma_T$ for the $T$-dependence of the longitudinal mode gap in 
the ordered phase remains a current theoretical challenge. For this it is 
important to recall that at finite temperature this Higgs mode becomes 
weakly overdamped, and what we show is the associated maximum in scattering 
intensity. For $p = 1.75$ kbar we find $\gamma_T = 0.54(8)$. 

Further insight into thermal scaling exponents can be obtained from the 
staggered magnetic moment, $m_s$, measured in Ref.~\cite{rrfsskgm}. Fits to 
the form $m_s \propto (T_N - T)^{\beta'}$ yield values close to the classical 
field-theory expectation \cite{rzj} $\beta'= 0.34$ (XY) or 0.37 [SU(2)]. 
However, the fit is followed only in a narrow window $T_{SL}(p) < T \le T_N(p)$, 
with $T_{SL} \simeq 0.8 T_N$ and points further away from the transition 
diverging clearly from classical scaling \cite{rrfsskgm}. Thus $m_s$ 
represents very well the scaling relations expected for a narrow CC regime 
in a system dominated by a QCP [Fig.~4(c)]. The narrow nature of the CC 
region is also evident in dynamical properties, in that the divergence of 
the QC and longitudinal mode widths may be used to set dynamical criteria 
for the crossover from QC to CC scaling. On the right side of Figs.~3(c), 
$\alpha_L$ approaches a constant also at finite $T$, but diverges on 
approaching $p_c(T)$ (on the left side $\alpha \rightarrow 0$ towards the 
QD regime). In Fig.~3(d), quantum behaviour is evident in the constant 
values of $\alpha$ reached far from the critical points, a regime 
including the QC excitations around the line $p = p_c$ (where $\alpha = 
\alpha_c$). We define classical scaling when $\alpha > 1/2$, giving the 
shaded regions shown in Figs.~3(c) and (d).The largely symmetric form of 
the $\Gamma_{\bf Q}/\epsilon_{\bf Q}$ curves is the best available indicator 
for a classical scaling regime on the disordered side of $T_N(p)$. These 
parallel static and dynamic approaches indicate the width of the CC regime 
in both pressure and temperature [Fig.~4(c)].  

Our precise level of control over quantum criticality in TlCuCl$_3$ has 
inspired recent numerical and analytical studies of the finite-$T$ properties 
of dimer systems at the coupling-induced QPT. The authors of Ref.~\cite{rjs} 
argue that $m_s(p)$ and $T_N(p)$ should have the same behaviour, and 
demonstrate good scaling of $m_s$ with $T_N$ by Quantum Monte Carlo 
simulations, independent of the functional form in $p$; this technique 
cannot address the longitudinal mode. In an effective quantum field theory 
approach \cite{roks}, $m_s(p) \propto T_N(p) \propto \sqrt{p - p_c}$ for 
a linear $p$-dependence of the exchange couplings, and $\Delta_L (p)$ also 
has this dependence. This analysis is predicated on proximity to a QCP, 
but in neglecting the classical critical regime, the field theory does 
not return the correct behaviour for static quantities around $T_N$, 
although its dynamical predictions remain valid. The exchange anisotropy 
in TlCuCl$_3$ is found \cite{roks} to have small quantitative effects on 
the calculated quantities, but no detectable qualitative ones ({\it e.g.}~on 
exponents). From our measurements, the best fits to the pressure exponents 
for $m_s$ and $T_N$ lie close to the classical value of 0.35 \cite{rrfsskgm}, 
although the quantum value of 0.5 is not beyond the error bars very close to 
the QCP. From experiment, the two quantities scale well together near the QCP, 
as shown in Figs.~4(d) and (e), but depart from universal scaling \cite{rjs} 
around an ordered moment of 0.4$\mu_{\rm B}/{\rm Cu}$ [Fig.~4(e)].

We have shown that the effects on the spectrum of quantum and thermal 
melting are qualitatively very similar. Both result in the systematic 
evolution of excitations whose gap increases away from the classical phase 
transition line, rather than simply a loss of coherence due to thermal 
fluctuations. Microscopically, quantum fluctuations in a dimer-based system 
cause enhanced singlet formation and loss of interdimer magnetic correlations, 
while thermal fluctuations act to suppress the spin correlation function 
$\langle {\bf S}_i \cdot {\bf S}_j \rangle$ on both the dimer and interdimer 
bonds. These correlation functions may be estimated from neutron-scattering 
intensities \cite{rnru} and also measured in dimerized optical lattices of 
ultracold fermions \cite{rgujte}. In TlCuCl$_3$, both methods of destroying 
interdimer coherence cause the triplet modes to evolve in the same way. A 
key question in the understanding of quantum criticality is whether quantum 
and thermal fluctuations can be considered as truly independent, and whether 
this independence may be taken as a definition of the QC regime \cite{rjs}. 
Our experimental results suggest that weak departures from universality 
become detectable at $(p,T)$ values away from the QC and CC regimes, and 
particularly as we increase the excitation energy, presumably as microscopic 
details of the fluctuation redistribution cause a mixing of quantum and 
thermal effects.

Finally, the existence of the longitudinal ``Higgs'' mode has been questioned 
in the past. Its visibility has recently been analyzed in detail in the 
scaling limit \cite{rpaa,rps} for systems in two and three dimensions. Our 
results confirm that it is a genuine example of quantum critical dynamics 
in three dimensions. Its critical nature makes it infinitely susceptible to 
thermal fluctuations, so that it becomes overdamped as soon as these become 
noticeable. While still possessing a significant spectral weight, the 
longitudinal mode of the pressure-ordered phase is overdamped at finite 
temperatures, although its critical nature is restored on passing into the 
finite-$T$ disordered phase. There are several reasons for the ready 
visibility -- in the longitudinal rather than the scalar susceptibility 
\cite{rpaa} -- of the longitudinal mode in TlCuCl$_3$, meaning for the 
anomalously low value of $\alpha_L$, despite its critically damped nature. 
These include the high dimensionality of the system, its low phase space 
for magnon scattering, the collinearity of the ordered moments, and the fact 
that one of the spin waves contributing to decay processes is massive. It 
is reasonable to assume that the same factors also control the anomalously 
low value of $\alpha_c$, allowing ready observation of the QC excitations 
at $p = p_c$ [Figs.~4(a) and (b)], and the very narrow regime of logarithmic 
corrections to scaling \cite{clc}, which are completely absent in our data. 
Both the visibility of the longitudinal mode and our level of control over 
both quantum and thermal fluctuations in TlCuCl$_3$ remain significantly 
superiour to any other magnetic \cite{raw,rltn}, charge-density-wave 
\cite{rlv}, or cold-atom \cite{reea} systems displaying this ``Higgs boson.''

In summary, high-resolution neutron spectroscopy experiments on the quantum 
antiferromagnet TlCuCl$_3$ allow us to probe the spin excitations of all 
phases in and around the QC regime by varying the pressure and temperature. 
We demonstrate a number of remarkable properties arising at the interface 
between quantum and classical physics. Quantum and thermal fluctuations have 
remarkably similar effects in melting the magnetically ordered phase and in 
opening excitation gaps, but operate quite independently close to the QCP. 
In the QC regime there is robust $\omega/T$ scaling of the energies and 
widths of critically damped excitations. This scaling crosses over to a 
classical critical form in a narrow region around the phase transition 
line $T_N(p)$. The critically damped longitudinal, or Higgs, mode of the 
ordered phase is exquisitely sensitive to thermal fluctuations and becomes 
overdamped in the classical regime. 

\bigskip
\bigskip
\noindent
{\bf Materials and Methods}

High-quality single crystals of TlCuCl$_3$ were grown by the Bridgman 
method. INS studies were performed on the cold-neutron triple-axis 
spectrometer IN14 at the Institut Laue Langevin (ILL). This was operated 
at constant final wavevector $k_f = 1.15$ \AA$^{-1}$, with a focusing 
pyrolytic graphite analyser and monochromator, collimation 
open-60'-open-open and a cooled Be filter positioned between sample 
and analyser. The temperature and pressure were controlled with a 
He cryostat and a He gas pressure cell (precision $\pm 50$ bar). We 
fitted our experimental measurements with a thermal damping ansatz 
\cite{DHOHe}, which has been used for the accurate modelling of 
phonon damping at finite temperatures \cite{phononDHO} and shown in 
Ref.~\cite{rrmnmfkgbsm} to be reliable for triplet spin excitations. 
The magnon is modelled as a damped harmonic oscillator (DHO), whose 
scattering intensity has the double-Lorentzian lineshape
\begin{equation}
S({\bf Q},\omega) = \frac{A [n(\omega) + 1] 4 \Gamma_{\bf Q} \epsilon_{\bf Q} 
\omega}{[\omega^2 - \epsilon_{\rm DHO}({\bf Q})^2]^2 + 4 \Gamma_{\bf Q}^2 \omega^2},
\end{equation}
where $n(\omega)$ expresses the thermal magnon population. Here $\epsilon_{\rm 
DHO} ({\bf Q})^2 = \epsilon_{\bf Q}^2 + \Gamma_{\bf Q}^2$ is a renormalized energy 
expressed in terms of the real excitation energy, $\epsilon_{\bf Q}$, and the 
linewidth of the scattered intensity, taken as the FWHM, $\Gamma_{\bf Q}$. 
The fits presented in Fig.~2 are based on a four-dimensional convolution in 
momentum and energy of the model cross-section [Eq.~(1)] with the instrument 
resolution, which causes the asymmetric peak shapes. Excitations measured 
throughout the $(p,T)$ phase diagram were characterized in this way by their 
energies $\epsilon_{\bf Q}$, linewidths $\Gamma_{\bf Q}$, and intensities 
[Eq.~(1)]. 

%\noindent
%{\bf References}

\bigskip
\noindent
{\bf Acknowledgements}

\noindent
We are grateful to M. Vojta for invaluable discussions. We thank the sample 
environment team at the ILL, where these measurements were performed, for 
their assistance. This work was supported by the EPSRC, the Royal 
Society, the Swiss NSF, and the NSF of China under Grant No.~11174365.

\bigskip
\noindent
{\bf Author contributions}

\noindent
P.M. and Ch.R. carried out the experiments with the help of instrument 
scientist M.B. TlCuCl$_3$ single crystals were synthesized by K.W.K. 
The theoretical and experimental framework was conceived by Ch.R., D.F.M. 
and B.N. Data refinement and figure preparation were performed by P.M. 
and Ch.R. The text was written by B.N. and Ch.R. 

\bigskip
\noindent
{\bf Additional information}

\noindent
Correspondence and requests for information should be addressed to Ch.R. 
(christian.rueegg@psi.ch).

\bigskip
\noindent
{\bf Competing Financial Interests}

\noindent
The authors declare no competing financial interests.


\begin{thebibliography}{00}
\bibitem{clc} Logarithmic corrections are expected to be relevant in a 
system at the upper critical dimension ($d_c = 4$, as here \cite{raw}) as 
$p_c$ is approached \cite{rks}. However, the width of the ``log-correction'' 
regime is a non-universal quantity. Because our data provide no signs of such 
corrections in either the thermal or the pressure exponents, we conclude that 
this regime is unusually small in TlCuCl$_3$.
\end{thebibliography}
\end{document}